\documentclass[11pt,a4paper]{article}
\usepackage[T1]{fontenc}
\usepackage{mathptmx}
\usepackage[margin=1in]{geometry}
\usepackage{cite}
\usepackage{amsmath,amssymb}
\usepackage{graphicx}
\usepackage{booktabs}
\usepackage{adjustbox}
\usepackage{caption}
\usepackage{microtype}
\usepackage{fancyhdr}
\usepackage{placeins}
\graphicspath{{figures/}}
\begin{document}
\begin{center}
{\LARGE\bfseries PARIS: Parallel Arithmetic Reduction and Integrated Selection for MSDF Inner Products\par}
\vspace{1.2em}
{\large Malik Zohaib Nisar$^{1}$, Ic-Pyo Hong$^{1}$, Dorit Merhof$^{2}$, Muhammad Usman$^{2}$\par}
\vspace{0.65em}
{\small $^{1}$Department of Smart Information Technology Engineering, Kongju National University, Cheonan, Republic of Korea\par}
{\small $^{2}$Faculty of Informatics and Data Science, University of Regensburg, Germany\par}
\vspace{0.45em}
{\small \texttt{\{zohaib,iphong\}@kongju.ac.kr}; \texttt{\{muhammad.usman,dorit.merhof\}@ur.de}\par}
\vspace{0.45em}
{\small Corresponding author: Muhammad Usman (\texttt{muhammad.usman@ur.de})\par}
{\small This work was supported by the German Research Foundation (DFG), project number 573796083.\par}
\end{center}
\vspace{0.8em}
\begin{abstract}
Existing most-significant-digit-first (MSDF) accelerators serialize input-term processing, causing inner-product latency to increase with fan-in. This paper presents PARIS, an MSDF architecture that eliminates input-term serialization through parallel bit-plane reduction and shared carry-save residual selection. For $B$-bit operands, PARIS processes $B^2$ bit-position pairs, independent of fan-in within the supported tile, while generating exact MSDF outputs. Synthesized in 45\,nm at 100\,MHz, the 784-lane INT8 core occupies 0.008701\,mm$^2$ and completes an inner product in 88 cycles. Evaluation on two MNIST multilayer perceptrons demonstrates arithmetic-workload speedups of 240.44$\times$ and 180.47$\times$ over a matched reused-MSDF implementation, reducing arithmetic energy by 98.89\% and 98.58\%, respectively. A synthesized 16-engine accelerator occupies 0.90273\,mm$^2$ and meets the 100\,MHz timing constraint. Assuming immediate weight availability, it delivers a projected throughput of 24.33\,GOPS and an estimated energy efficiency of 1,581.14\,GOPS/W at unchanged mapped power. Bias-aware ReLU detection further reduces modeled arithmetic cycles by 5.52\% and 4.86\%. A separate 256-engine RTL implementation completes Model A in 259 cycles, demonstrating output-parallel scalability.
\end{abstract}
\vspace{0.8em}

\section{Introduction}
Fully connected layers require inner products over many input terms, contributing significantly to inference latency and energy consumption. Bit-parallel arithmetic provides high throughput but requires more hardware resources, whereas bit-serial arithmetic reduces hardware complexity at the cost of longer execution time. Consequently, reducing datapath width does not necessarily improve energy efficiency.

Most-significant-digit-first (MSDF) arithmetic generates the most significant result digits first, allowing successive operations to overlap after an initial delay $\delta$ \cite{ercegovac2004digital}. Recent MSDF accelerators reuse multiply--accumulate (MAC) units across the input terms of each neuron \cite{cherati2025mac,cherati2025termination,usman2025energy}. Although neurons can operate in parallel, their input terms are processed sequentially, causing inner-product latency to increase with fan-in. Other bit-serial accelerators improve throughput through reduced operand precision \cite{judd2016stripes,sharify2018loom,umuroglu2019bismo}. These approaches motivate an MSDF architecture that processes input terms concurrently while retaining digit-serial computation.

This paper presents PARIS (Parallel Arithmetic Reduction and Integrated Selection), an MSDF architecture that processes input terms in parallel within each tile while evaluating operand bit-position pairs sequentially. A shared carry-save residual datapath integrates digit selection and residual correction to generate exact MSDF outputs. Unlike activation-only bit-serial designs with parallel weight bits, PARIS processes both operand bit positions sequentially. For $B$-bit operands, it requires $B^2$ bit-position steps, independent of fan-in within the supported tile. The main contributions are:

\begin{itemize}
\setlength{\itemsep}{1pt}\setlength{\parsep}{0pt}
\item An aggregate MSDF datapath combining parallel bitwise operations, population-count reduction, and shared carry-save residual selection to compute exact inner products without serializing input terms within a tile.
\item A bias-aware ReLU certificate that identifies nonpositive pre-activations early, with bit-exact verification and modeled arithmetic-cycle savings.
\item A 45-nm ASIC evaluation covering arithmetic-core performance, fan-in and precision scaling, and a synthesized 16-engine accelerator, complemented by a 256-engine RTL implementation demonstrating output-parallel scalability.
\end{itemize}

The 784-lane INT8 core completes an exact inner product in 88 cycles. The evaluation includes matched arithmetic-core comparisons and accelerator-level analysis under a resident-weight operating assumption.

\section{Proposed PARIS Architecture}
\label{sec:architecture}

\subsection{MSDF Representation and Parallel Bit-Plane Reduction}

\begin{figure}[!ht]
    \centering
    \includegraphics[width=0.87\textwidth]{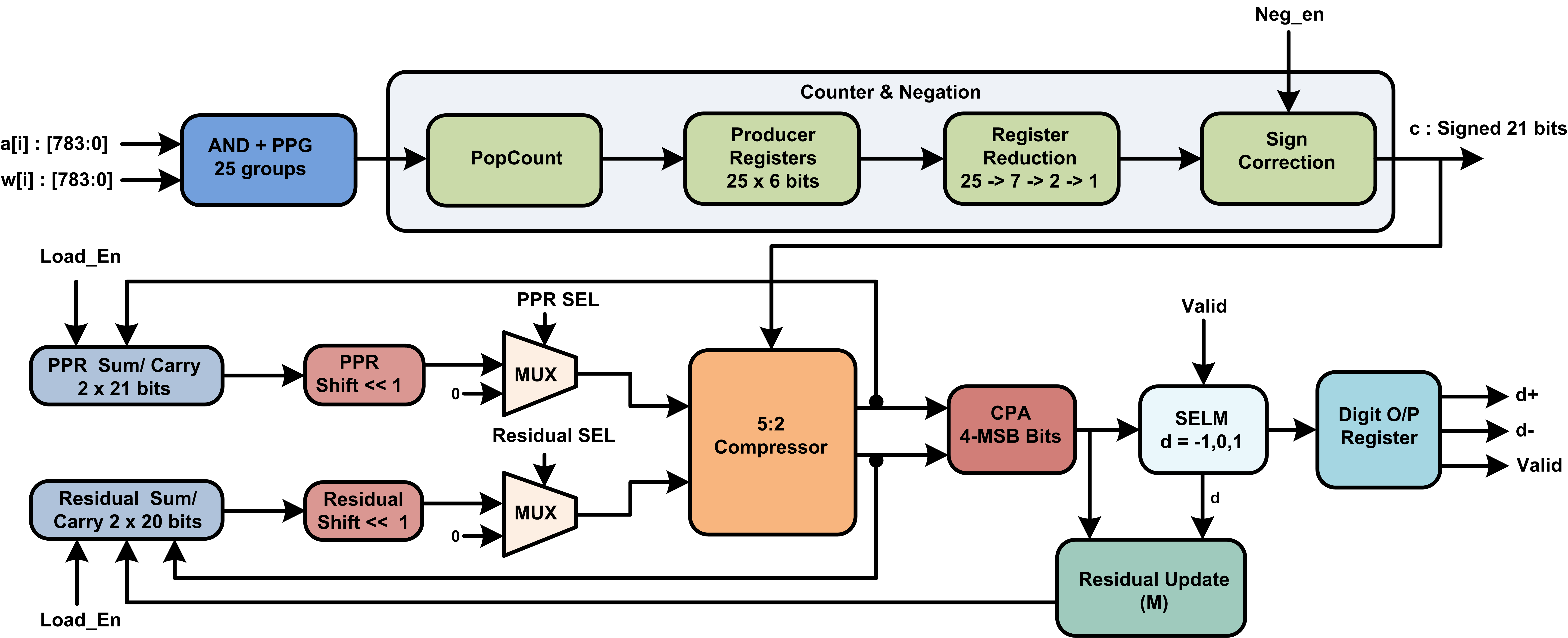}
    \caption{Architecture of the proposed PARIS inner-product unit.}
    \label{fig:datapath}
\end{figure}

PARIS computes exact inner products from signed INT8 two's-complement operands and generates MSDF output digits $d\in\{-1,0,+1\}$. Each digit uses two-bit encoding $(d^+,d^-)$, representing $+1$, $-1$, and zero as $10$, $01$, and $00$, respectively. The output position advances only when the digit-valid signal is asserted.

Fig.~\ref{fig:datapath} shows the architecture with $L=784$ parallel lanes. For activation and weight bit planes $a[i]$ and $w[j]$, parallel AND operations and population counting produce $c_{i,j}=\operatorname{popcount}(a[i]\land w[j])$. The signed inner product is expressed as
\begin{equation}
S=\sum_{n=0}^{L-1}a_nw_n
=\sum_{i,j=0}^{B-1}\sigma_i\sigma_j2^{i+j}c_{i,j},
\label{eq:bitplane}
\end{equation}
where $\sigma_{B-1}=-1$ and $\sigma_i=+1$ otherwise account for the two's-complement sign bits. PARIS processes all input terms in $B^2$ bit-position steps (64 for INT8). Sign correction negates the count when exactly one selected position is a sign bit.

The partial-product generation (PPG) unit divides 784 lanes into 24 groups of 32 and one of 16. Six-bit local counts pass through three registered reduction stages ($25\rightarrow7\rightarrow2\rightarrow1$) using 8-, 10-, and 10-bit sums. The final ten-bit count ranges from 0 to 784. Control and valid signals are pipelined with the data.

\subsection{Carry-Save Residual and Shared Selection}

PARIS processes activation and weight bits from $B-1$ to zero. For each activation bit position $i$, the partial-product row (PPR) is constructed by shifting the accumulated value left and adding successive signed population counts. The completed row and inner-product recurrence are

\begin{equation}
q_i=\sigma_i\sum_{n=0}^{L-1}a_n[i]w_n,\qquad
S_r=2S_{r-1}+q_i,
\label{eq:row_recurrence}
\end{equation}

where $i=B-1-r$ and $S_{-1}=0$. Multiplication by two requires only a one-bit shift.

The PPR and residual are maintained in separate carry-save sum/carry register pairs. A shared 5:2 compressor combines the signed population count with the shifted register values, avoiding full-width carry propagation. Feedback multiplexers clear the PPR at the start of each row, accumulate intermediate counts, and combine the completed row with twice the residual at the final weight bit. During residual draining, the count and PPR contributions are set to zero.

At each row completion, the selected MSDF digit is subtracted to update the residual:

\begin{equation}
V_r=2R_{r-1}+q_i,\qquad
R_r=V_r-Hd_r,\quad R_{-1}=0,
\label{eq:residual}
\end{equation}

where $H=2^{B+G+1}$, $G=\lceil\log_2 L\rceil$, and $u=H/4$. Integrated selection combines digit selection and residual correction within the shared carry-save datapath (Fig.~\ref{fig:datapath}). A four-bit carry-propagate adder (CPA) estimates $V_r$ from the leading carry-save bits as $V_r=hu+\ell_0+\ell_1$, where $0\le\ell_0,\ell_1<u$. Since $|q_i|\le u$, the residual bound $-2u\le R<3u$ restricts $h$ to $[-6,6]$. The selector emits $d=-1$ for $h\le-3$, $d=0$ for $-2\le h\le1$, and $d=+1$ for $h\ge2$, preserving these bounds.

For nonzero digits, the Residual Update (M) block applies the correction by toggling the $H$ bit, as adding or subtracting $H$ is equivalent modulo $2H$. This avoids full-width subtraction and allows a single constant-width digit selector to serve the entire inner product.

The emitted digits are accumulated as $D_r=2D_{r-1}+d_r$, preserving the invariant

\begin{equation}
S_r=HD_r+R_r.
\label{eq:reconstruction}
\end{equation}

After processing all rows, $J=\log_2H$ residual-drain cycles complete digit generation. The invariant then becomes $HS=HD+R$. Since $S$ and $D$ are integers, $R$ must be a multiple of $H$. The maintained bound $-H/2\le R<3H/4$ therefore requires $R=0$, giving $D=S$. Suppressing the initial zero leaves $2B+G=26$ exact signed digits for the INT8, 784-lane implementation.

\subsection{Schedule, Tiling, and Operand Delivery}

PARIS processes all 64 INT8 bit-position pairs in consecutive cycles. Five pipeline cycles and 19 residual-drain cycles give an inner-product latency of 88 cycles. After suppressing the initial zero, valid output digits appear at cycles 21, 29, $\ldots$, 69, followed by 19 consecutive drain cycles. With one restart cycle, the latency and initiation interval are
\begin{equation}
C_{\rm dot}=B^2+5+(B+G+1)=88,\qquad
I_{\rm dot}=89.
\label{eq:cycles}
\end{equation}
For fixed $B$ and $L$, latency is independent of operand values and active fan-in $N\le L$.

For a network with $N_\ell$ inputs, $M_\ell$ neurons, and $T_m$ parallel output engines per layer, the scheduled tile-group count is
\begin{equation}
Q=\sum_\ell
\left\lceil\frac{M_\ell}{T_m}\right\rceil
\left\lceil\frac{N_\ell}{L}\right\rceil.
\label{eq:tiles}
\end{equation}
The corresponding core-only execution time is
\begin{equation}
C_{\rm net}=(Q-1)I_{\rm dot}+C_{\rm dot},
\label{eq:network}
\end{equation}
assuming continuous operand delivery and excluding cross-tile accumulation and layer-transition overhead. Increasing the number of output engines or input lanes reduces the required tile groups.

The core receives activation and weight bit planes each cycle, with operands reused across 64 bit-position pairs. Evaluation uses a preloaded behavioral operand producer, excluding storage and transfer costs from the reported core results.

\begin{table}[t]
\caption{Matched MSDF arithmetic cores at 100 MHz.}
\label{tab:matched}
\centering
\footnotesize
\setlength{\tabcolsep}{3pt}
\begin{tabular}{lrrrr}
\toprule
& \multicolumn{2}{c}{Reused MSDF}
& \multicolumn{2}{c}{\textbf{PARIS (ours)}}\\
\cmidrule(lr){2-3}\cmidrule(lr){4-5}
Metric & Model A & Model B & Model A & Model B\\
\midrule
Cycles
& 5,691,923 & 4,979,019 & 23,673 & 27,589\\
Latency ($\mu$s)
& 56,919.23 & 49,790.19 & 236.73 & 275.89\\
Throughput (GOPS)
& 0.0071 & 0.0071 & 1.7172 & 1.2889\\
Core cell area (mm$^2$)
& \multicolumn{2}{c}{0.002479}
& \multicolumn{2}{c}{0.008701}\\
Power (mW)
& 0.1415 & 0.1414 & 0.3778 & 0.3620\\
Energy ($\mu$J)
& 8.0550 & 7.0401 & 0.0895 & 0.0999\\
GOPS/W
& 50.47 & 50.51 & 4,544.74 & 3,560.48\\
EDP ($\mu$J$\cdot\mu$s)
& 458,483.76 & 350,529.42 & 21.18 & 27.55\\
GOPS/mm$^2$
& 2.88 & 2.88 & 197.35 & 148.13\\
Speedup
& 1.00 & 1.00 & 240.44 & 180.47\\
\bottomrule
\end{tabular}
\end{table}

\subsection{ASIC Implementation}

PARIS and the reference design were synthesized using Synopsys Design Compiler with the Nangate 45-nm library at 1.1\,V, and 100\,MHz. Both designs have zero setup, hold, and design-rule violations, with setup slacks of 3.71\,ns and 3.95\,ns, respectively. PARIS occupies 0.008701\,mm$^2$ with 4,981 cells and 349 flip-flops, compared with 0.002479\,mm$^2$, 1,367 cells, and 210 flip-flops for the reference. Area represents the sum of mapped standard cells.

In PARIS, combinational and sequential logic account for 81.83\% and 18.17\% of the area. The producer/interface, global reduction, compressor, and residual/selection/control blocks occupy 67.99\%, 15.76\%, 4.35\%, and 11.90\%, respectively. Their corresponding dense-workload power shares are 69.36\%, 16.04\%, 2.48\%, and 12.12\%. Internal, switching, and leakage power are 0.5496, 0.1648, and 0.0947\,mW, respectively. Power is estimated from mapped gate-level activity using four images per model and library leakage, excluding physical implementation effects.

\subsection{MLP Workloads}

We evaluate MNIST~\cite{lecun1998gradient} Models A (784--256--10) and B (784--200--100--10), matching the topologies in~\cite{cherati2025mac} but independently trained for 200 epochs using 50,000 training and 10,000 validation images. Calibration uses 2,048 training images. INT8 accuracy on the separate 10,000-image test set is 98.04\%/98.12\%.

Pixels are quantized as $\operatorname{RNE}(127p/255)$, weights use a per-layer scale of $\max|W|/127$, and INT32 biases use the product of activation and weight scales. Hidden layers apply ReLU, calibrated Q31 multiply/shift with nearest-even rounding, and saturation to 0--127. Final classification uses argmax of biased sums with first-index tie breaking. Accuracy is evaluated in integer software, independently of hardware timing.

\subsection{Matched MSDF Baseline and Evaluation Metrics}

For a matched comparison, we implemented a reused-MSDF MAC based on the architectures in \cite{cherati2025mac}. The baseline uses a serial--serial multiplier and redundant accumulator and is an independent architectural reimplementation rather than the authors' original RTL. Both designs use a single output engine, signed INT8 operands, the same 45-nm standard-cell library, and a 100\,MHz clock. Both produce exact 26-digit inner products.

In the baseline, the multiplier generates 16 product digits that are aligned with a 26-digit redundant accumulator. Ten leading positions accommodate the multiplier start-up, while its three-cycle delay is overlapped with accumulation. Each multiplication requires 26 digit-processing cycles and two drain cycles. For an inner product with $N$ active terms, the resulting latency is $28N+1$ cycles, with an initiation interval of $28N+2$ cycles. No additional idle cycles are introduced between terms.

Models A and B require 406,528 and 355,600 useful operations per inference, respectively, counting multiplication and addition as two operations per MAC. Zero-padding operations are excluded. Latency is measured from the first start to the final valid output digit. The execution interval is $C_{\rm int}=C_{\rm lat}+1$, giving throughput $\mathcal{T}=Of/C_{\rm int}$ and energy $E=PC_{\rm int}/f$, where $O$, $P$, and $f$ denote the operation count, average power, and clock frequency. Energy efficiency is $\mathcal{T}/P$, while the energy--delay product (EDP) is $EC_{\rm lat}/f$.

Bit-exact verification covers signed operand extremes, random inputs, and active fan-ins from 1 to 784. Gate-level switching activity is evaluated using 1,064 and 1,240 inner products for Models A and B, respectively. The corresponding MNIST activation traces contain 82.3\% and 77.7\% zero activation terms, allowing power estimation under representative workloads. The matched evaluation isolates arithmetic-core performance under identical precision, technology, and clock conditions, enabling a direct comparison of the two architectures. Network accuracy is evaluated independently using integer inference.

\section{Results and Discussion}

\subsection{Performance and Energy Efficiency}

Six independently synthesized configurations passed bit-exact RTL and mapped-gate verification. At INT8 precision, increasing fan-in from 256 to 1024 changes dot-product latency by only two cycles (86 to 88), while area and power increase with lane count. At 784 lanes, reducing precision to INT6 and INT4 lowers latency from 88 to 58 and 36 cycles, respectively, with corresponding power reductions.

Energy per dot increases with fan-in, but energy per MAC decreases as shared overhead is amortized across more lanes. At 784 INT8 lanes, dense-input operation requires 0.92\,pJ/MAC and consumes 2.14$\times$ the power of the MNIST workload, highlighting the influence of switching activity.
\begin{figure}[h]
 \centering
 \includegraphics[width=0.5\columnwidth]{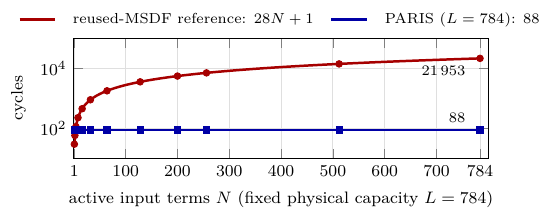}
 \caption{Complete-dot cycles versus active fan-in.}
 \label{fig:fanin}
\end{figure}

At 100\,MHz, PARIS achieves 240.44$\times$ and 180.47$\times$ speedups over the matched reused-MSDF core for Models A and B, respectively (Table~\ref{tab:matched}). Its dot-product latency remains nearly independent of fan-in, whereas the reference latency increases linearly (Fig.~\ref{fig:fanin}). Despite requiring 3.51$\times$ the cell area and approximately 2.6$\times$ the power, PARIS reduces arithmetic energy by 98.89\% and 98.58\% for the two models.
\begin{figure}[h]
 \centering
 \includegraphics[width=0.45\columnwidth]{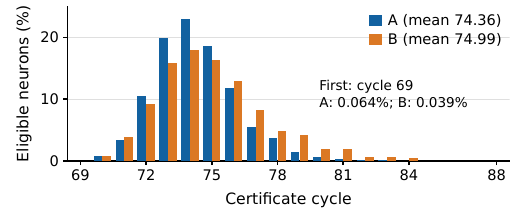}
 \caption{Certificate cycles, conditional on nonpositive biased hidden preactivation; 10,000 images per model.}
 \label{fig:exit}
\end{figure}

The MSDF digit stream also enables exact bias-aware ReLU termination. After $k$ of 26 digits, the bound $U_k=P_k+b+(2^{26-k}-1)\le0$ certifies a zero ReLU output and permits early restart. Evaluation on 10,000 test images predicts 5.52\% and 4.86\% fewer serialized arithmetic cycles for Models A and B, respectively (Fig.~\ref{fig:exit}), without changing predictions. RTL and mapped-core checks confirm correct prefix evaluation and restart behavior. These savings represent modeled arithmetic-cycle reductions rather than measured energy savings.
\label{sec:results}

\subsection{Comparison with Existing Accelerators}

Table~\ref{tab:published} compares PARIS with published MLP accelerators. For the 784--256--10 network, the 16-engine resident-weight configuration delivers a projected 24.33\,GOPS at 100\,MHz. Using the synthesized power estimate of 15.39\,mW, its estimated energy efficiency reaches 1,581.14\,GOPS/W, numerically 3.46$\times$ that of the MSDF MAC~\cite{cherati2025mac}, assuming unchanged average power.

The integrated 16-engine accelerator was synthesized in 45-nm technology, including weight storage, arithmetic engines, requantization, and control. It occupies 0.90273\,mm$^2$ and meets the 100\,MHz timing constraint with 3.96\,ns positive setup slack. A separate 256-engine RTL implementation completes Model A in 259 cycles, corresponding to 156.36\,GOPS if operated at 100\,MHz. Bit-exact RTL verification across 128 test images confirms its functional correctness.
\begin{table}[t]
\caption{Comparison of existing MLP accelerators and the proposed PARIS architecture.}
\label{tab:published}
\centering
\footnotesize
\setlength{\tabcolsep}{1.5pt}
\renewcommand{\arraystretch}{0.95}
\begin{adjustbox}{max width=\textwidth}
\begin{tabular}{l*{10}{c}}
\toprule
& \shortstack{MSDF MAC\\ \cite{cherati2025mac}}
& \shortstack{MSDF MLP (12/12)\\\cite{cherati2025termination}}
& \shortstack{MSDF MLP (6/6)\\\cite{cherati2025termination}}
& \shortstack{Kim\\ \cite{kim2017}}
& \shortstack{Tu\\ \cite{tu2019}}
& \shortstack{Westby\\ \cite{westby2021}}
& \shortstack{R. Wang\\ \cite{wang2017}}
& \shortstack{Sarwar\\ \cite{sarwar2018}}
& \shortstack{Y. Wang\\ \cite{wang2021}}
& \shortstack{\textbf{PARIS }\\16 engines$^\dagger$}\\
\midrule

Model
& \shortstack{784--256\\--10}
& \shortstack{784--256\\--10}
& \shortstack{784--256\\--10}
& \shortstack{784--144\\--10}
& \shortstack{784--256\\--10}
& \shortstack{784--12\\--10}
& \shortstack{784--8,000\\--10}
& \shortstack{784--1,200\\--600--10}
& \shortstack{784--1,200\\--1,200--10}
& \shortstack{784--256\\--10}\\

Technology
& 45 nm & 45 nm & 45 nm
& 130 nm & 65 nm
& \shortstack{FPGA\\16 nm}
& \shortstack{FPGA\\28 nm}
& 45 nm
& \shortstack{FPGA\\20 nm}
& 45 nm\\

Clock (MHz)
& 588 & --- & ---
& 200 & 500 & 100 & 266 & --- & 250
& 100\\

Bits
& 8 & 12 & 12
& 8 & 16 & 32 & 18 & 12 & 13
& 8\\
\midrule

Accuracy (\%)
& 96.73 & 96.57 & 95.13
& 94.00 & 96.40 & 93.25 & 96.55
& 98.88 & 98.44 & 98.04\\

Latency ($\mu$s)
& 1.46 & 2.12 & 2.12
& 2.98 & 25.63 & 1.55 & 120
& --- & 845 & 16.70\\

Power (mW)
& 611 & 839.80 & 477.59
& 2,151 & 131.87 & 567.74 & ---
& --- & 441.90 & 15.39\\

Area (mm$^2$)
& 5.88 & --- & ---
& 6.76 & 0.55 & --- & ---
& --- & --- & 0.90273\\

Energy ($\mu$J)
& 0.89 & 1.780 & 1.012
& 6.41 & 3.38 & 0.88 & ---
& 1,820 & 373.41 & 0.2570\\

GOPS
& 278.44 & 191.76 & 191.76
& 76.74 & 15.86 & 12.29 & ---
& --- & --- & 24.33\\
\midrule

GOPS/W
& 456.77 & 228.39 & 401.71
& 35.67 & 120.27 & 21.65 & ---
& --- & --- & \textbf{1,581.14}\\

\bottomrule
\end{tabular}
\end{adjustbox}

\par\vspace{3pt}
\parbox{\textwidth}{\footnotesize
$^\dagger$PARIS uses a 16-engine resident-weight configuration. Performance is projected
using the synthesized 100\,MHz implementation and its mapped average power (15.39\,mW). Published results retain their reported implementation conditions.
}
\end{table}

\section{Conclusion}

This paper presented PARIS, an MSDF architecture that processes input terms in parallel within each tile through bit-plane reduction and shared residual selection. The 784-lane INT8 core computes exact inner products in 88 cycles, independent of active fan-in within the supported tile. Compared with a matched reused-MSDF implementation, PARIS achieves 240.44$\times$/180.47$\times$ speedups and reduces arithmetic energy by 98.89\%/98.58\% for Models A/B. Bias-aware ReLU termination further reduces modeled serialized-arithmetic cycles by 5.52\%/4.86\%. A 256-engine RTL implementation completes Model A in 259 cycles, demonstrating output-parallel scalability. Future work will address physical implementation and integrated operand storage.

\FloatBarrier


\end{document}